\documentclass{elsart}
\usepackage{epsf,amssymb}
\usepackage{graphicx}
\begin{document}
\begin{frontmatter}

\title{Ab initio approach for atomic relaxations  
in supported magnetic clusters}

\author[MPI]{V.S. Stepanyuk \corauthref{cor}},
\corauth[cor]{Corresponding author. E-mail: stepanyu@mpi-halle.de,
                                    Phone: +49 345 5525429,
                                    Fax: +49 345 5525446}
\author[UNI]{A.L. Klavsyuk},
\author[MPI]{L. Niebergall},
\author[MSU]{A.M. Saletsky},
\author[UNI]{W. Hergert},
\author[MPI]{P. Bruno}

\address[MPI]{Max-Planck-Institut f\"ur Mikrostrukturphysik, Weinberg
2, 06120 Halle, Germany }

\address[UNI]{Fachbereich Physik, Martin-Luther-Universit\"at,
Halle-Wittenberg, Friedemann-Bach-Platz 6,
 D-06099 Halle, Germany}

\address[MSU]{General Physics Department, Moscow State
University, 119899 Moscow, Russia}

\begin{abstract}
 
We present a newly developed scheme  for atomic relaxations of 
magnetic supported clusters.  
Our approach  is based on  
the full potential  Korringa-Kohn-Rostoker Green's function method  
and  the second moment tight-binding approximation for many-body 
potentials. 
We demonstrate that only a few iterations in ab initio calculations are 
necessary to 
 find an equilibrium structure of 
supported clusters. 
As an example, we present our results for small Co clusters on Cu(001). 
Changes in 
electronic and magnetic states  of  clusters due to atomic 
relaxations are revealed. \\

\end{abstract}
\end{frontmatter}

\newpage
\section{Introduction}
One of the central issues in physics of nanostructures is the interplay
between atomic structure, electronic  and magnetic states.
Different empirical and semi-empirical potentials  
have been used to determine the
interatomic interactions (Brenner, 2000).
These analytic potential energy functions can be considered as simplified
mathematical expressions for modeling interatomic forces arising from 
quantum mechanical interactions. Parameters of such  potentials are usually 
fitted to bulk properties which include lattice constants, cohesive 
energies, bulk modulus, elastic properties and vacancy formation energies. 
Although these   methods have given an 
important results on cluster formation, their interactions and diffusion
on metal surfaces,  they  become less effective for a 
quantitative description of structural relaxations in transition metal  
clusters on surfaces.   The lack of a detailed description of electronic 
states is also an obvious drawback of analytical potential energy 
functions. 
Several calculations of structural relaxations, electronic and magnetic 
properties  of small  magnetic clusters 
on metal surfaces have been recently  
performed using  a quantum-chemical methods(Nayak et al., 2001). A metal  
surfaces 
in such 
calculations is approximated  by the finite atomic cluster. One of the 
difficulties associated with using a cluster to model an infinite or 
semi-infinite system is that the cluster size is usually small due to 
computational limitations, and one must be sure that this limitation does 
not lead to spurious conclusions.   

Recently we have developed N-body interatomic potentials formulated in the 
second moment tight-binding approximation (TB-SMA) (Levanov et al., 2000) 
Parameters of 
potentials are fitted to   
accurate first-principle calculations of selected cluster-substrate 
properties. This scheme allows us to correctly reproduce bonding in 
transition metal clusters on metal surfaces. Several applications of our 
potentials can be found in  recent publications (Levanov et al., 2000; 
Stepanyuk et. al., 2000; Stepanyuk et al., 2003; Tsivlin et al., 2003, 
Longo et al., 2004).  
In this work we show that ab initio fitted many body potentials and the 
KKR Green's function method can be combined  to 
perform fully ab initio relaxations of magnetic clusters on an 
semi-infinite metal substrates. 
Many-body potentials are used to approximate trajectories of 
atoms during the relaxation.  Ab initio   
Helmmann-Feynman (HF) forces are calculated  to find an 
equilibrium structure.   
Structure, electronic and magnetic states 
are calculated  self-consistently. We concentrate on small Co 
clusters on Cu(100). 

\section{Method}
First, we describe the KKR Green's function method used for calculations  
of forces acting on atoms in supported clusters, and their electronic and 
magnetic properties (Wildberger et al., 1995; Papanikolaou et al., 1997).
The
basic idea of the method is a hierarchical scheme for the
construction of the Green's function of clusters on a metal surface
by means of successive applications of Dyson's equation. We treat
the surface as the two-dimensional perturbation of the
bulk. For the calculation of the ideal surface the nuclear
charges of several monolayers are removed, thus creating two half
crystals being practically uncoupled. Taking into account the 2D
periodicity of the ideal surface,  we find the structural
Green's function by solving a Dyson equation self-consistently:
\begin{eqnarray}
G_{LL'}^{jj'}({\bf k}_\parallel , E)=
\mbox{\r G}_{LL'}^{jj'}({\bf k}_\parallel , E)+
\sum\limits_{j''L''}
\mbox{\r G}_{LL''}^{jj''}({\bf k}_\parallel , E)
\Delta t_{L''}^{j''}(E)
G_{L''L'}^{j''j'}({\bf k}_\parallel , E).
\end{eqnarray}
Here $\mbox{{\r G}}$  is the structural Green's function of the bulk in a
${\bf k}_\parallel$-layer representation ($j,j'$ - layer indices).
The ${\bf k}_\parallel$ wave vector belongs to the 2D Brillouin
zone. $\Delta t_{L}^{j}(E)$ is the perturbation of the $t$ matrix
to angular momentum $L=(l,m)$ in the $j$-th layer.

The consideration of clusters on the surface destroys the
translation symmetry. Therefore the Green's function of the
cluster on the surface is calculated in a real
space formulation.
 The structural Green's function of the ideal surface  in real
space representation is then
used as the reference Green's function for the calculation of the
cluster-surface system  from an algebraic Dyson equation:
\begin{eqnarray}
G_{LL'}^{nn'}(E)=
\mbox{\r G}_{LL'}^{nn'}(E)+
\sum\limits_{n''L''}
\mbox{\r G}_{LL''}^{nn''}(E)
\Delta t_{L''}^{n''}(E)
G_{L''L'}^{n''n'}(E),
\end{eqnarray}
where $G_{LL'}^{nn'}(E)$  is the energy-dependent structural
Green's function matrix and $\mbox{\r G}_{LL''}^{nn''}(E)$ the
corresponding matrix for the ideal surface, serving as a reference
system.  $\Delta t_{L}^{n}(E)$ describes the difference in the
scattering properties at site $n$ induced by the existence of the
adsorbate atom.
Exchange and correlation effects are included in the local density
approximation. The full charge density and the full potential 
are used in our calculations.
For atomic relaxations of clusters we transform  the Green's function of 
the cluster to  the 
shifted coordinate system using the transformation matrix 
(Papanikolaou et al., 1997). 
In solving Schr\"odinger and Poisson equations we use the shape function.
For distorted atomic positions we make a Wigner-Seitz (WS) construction in the 
distorted geometry. We apply the ionic Hellmann-Feynman formula for 
calculations of forces acting on atoms(Papanikolaou et al., 1997).

Many-body potentials are formulated in the second moment tight 
binding approximation (Levanov et al., 2000).  
The attractive term (band energy) $E_{B}^{i}$  contains
the many-body interaction.
The repulsive term  $E_{R}^{i}$ is described by pair
interactions (Born-Mayer form).
The cohesive energy $E_{coh}$ is  the sum of band energy and repulsive 
part:
\begin{eqnarray}
E_{coh}   &=& \sum_{i}\left(E_{R}^{i}+E_{B}^{i}\right) \\
E_{R}^{i} &
=& \sum_{j}A_{\alpha\beta}
\exp(-p_{\alpha\beta}(\frac{r_{ij}}{r_{0}^{\alpha\beta}}-1)) \\
E_{B}^{i} &=
& -\left(\sum_{j}\xi_{\alpha\beta}^{2}\exp(-2q_{\alpha\beta}(\frac{r_{ij}}
{r_{0}^{\alpha\beta}}-1))\right)^{1/2}
\end{eqnarray}
$r_{ij}$ is the distance between
the atoms $i$ and $j$. $r_{0}^{\alpha\beta}$  is
the first neighbors distance
in the crystalline structures of the pure metals for
atom-like interactions and becomes
an adjustable parameter in the case of the cross interaction.
$\xi$ is an effective hopping integral; $p_{\alpha\beta}$ and
$q_{\alpha\beta}$
describe the decay of the
interaction strength with distance of the atoms.

Binding energies of supported  and embedded clusters of 
different sizes 
and geometry, the HF 
forces acting on adatoms and surface energies are accurately fitted to the 
ab initio results to 
find parameters of potentials.  
We should note that bonds of low coordinated atoms are 
considerably stronger than bonds in the bulk. Therefore, surface 
properties should be included in fitting of potentials to correctly 
describe atomic relaxations in supported clusters.
To link the interaction between atoms on the surface to that in the bulk 
the set of data used for fitting includes also such bulk properties as 
bulk modulus, lattice constant, cohesive energy, and elastic constants.
The combination of ab initio and TB methods allows  one to construct 
many-body potentials for low-dimensional structures and to perform atomic 
relaxations for very large systems. Parameters of potentials for Fe, Co 
and Cu nanostructures on Cu substrates can be found in our recent 
publications (Levanov et al., 2000; Longo et al., 2004; Stepanyuk et al., 
2004). 
Here, we show that these potentials can be used to perform a fully ab 
initio atomic relaxations by means of the KKR Green's function method.
The key idea of our approach is presented in Fig.1.
\begin{figure}[htp]
\begin{center}
   \includegraphics[width=14cm]{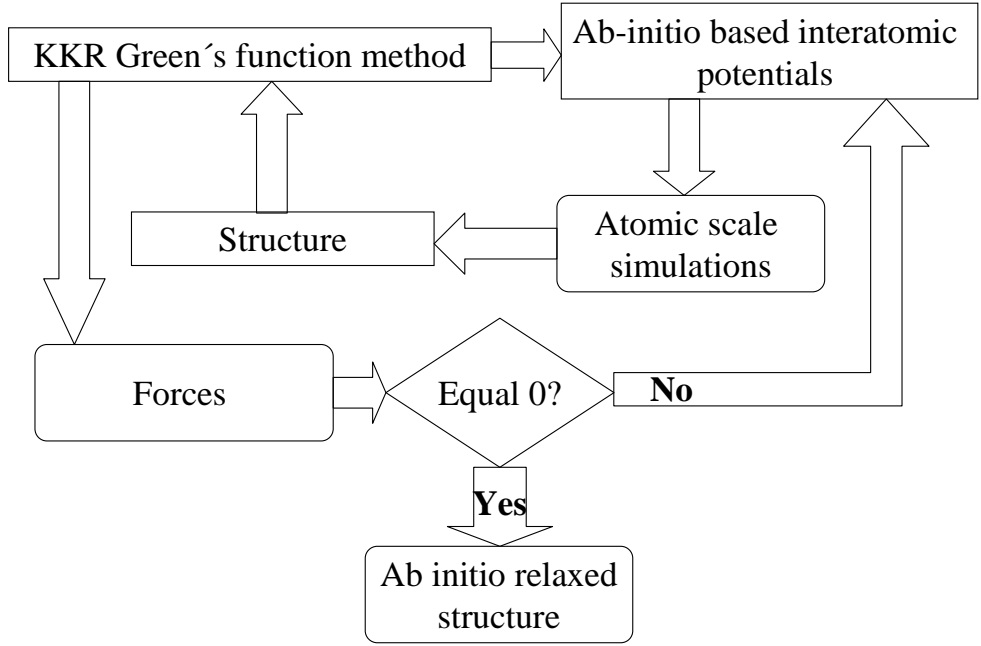}
\caption{Schematic presentation of the method for relaxations of supported
clusters; KKR Green's function method and ab initio fitted many-body
potentials are used to determine structure, electronic and magnetic
states in clusters fully self-consistently.}
\label{fig1}
\end{center}
\end{figure}
First, we find the  relaxed atomic configuration of the cluster on the 
surface   using  ab initio fitted many-body 
potentials. We call these potentials as the trial ones. 
This structure is used to perform the ab initio   
self-consistent calculations by means of the KKR Green's 
function method. Then, we calculate the HF forces acting on each atom.  
If these forces are not equal zero, the potentials are fitted again to 
correctly reproduce these ab initio forces. The structure of the cluster 
is again relaxed with a new fitted potentials. We repeat calculations until 
ab initio forces are very small, i.e. the cluster is fully relaxed. 
 We should note that, in fact, ab initio relaxations of supported clusters 
without using of potentials, are possible, however they require much more 
computational efforts. The problem is that any small changes in the 
position of some atom of the cluster during relaxation lead to strong 
changes in forces acting on different atoms.  Ab 
initio fitted many-body potentials allow us to find a very good 
approximation 
for  displacements of atoms at each iteration. In other words, potentials 
'help' the ab initio method  to find  correct atomic displacements.
We will show that only 6-7 iterations should be performed to fully relax 
small magnetic clusters.
Magnetic and electronic properties of clusters are determined 
self-consistently at each iteration.    

\section{Small Co clusters on Cu(001)}
First, we discuss in detail results for the Co$_4$ cluster.
As the first application of the method, we present calculations for  
atomic relaxations in 
clusters on the ideal Cu substrate. The ab initio relaxations of the 
substrate atoms can also be  performed  using our approach. However, 
as we will see 
in this section,  ab initio based potentials give very good approximation 
to the fully ab initio results. Therefore, we believe that reliable results 
for atomic relaxations in the substrate can be simply obtained 
using our potentials without performing a fully ab initio 
calculations (Stepanyuk et al., 2000; Tsivlin et al., 2003; Longo et al, 
2004;)

In Fig.2 we present the potentials for Co-Cu and Co-Co interactions 
obtained by fitting to ab initio results(the trial potentials).
\begin{figure}[htp]
\begin{center}
 \includegraphics[width=10cm]{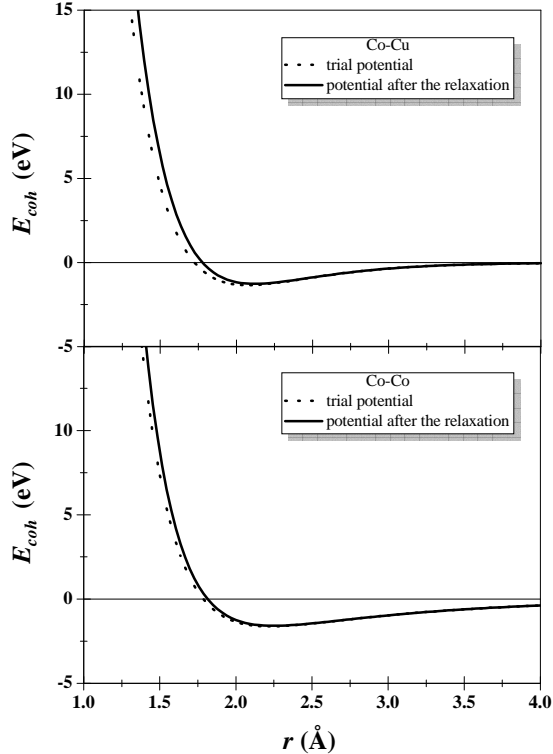}
\caption{The potentials for Co-Cu and Co-Co interactions.}\label{fig2}
\end{center}
\end{figure} 
Using these potentials we have carried out the relaxation of the 
structure of the Co$_4$  cluster. The bond length and the vertical 
coordinates of atoms before and after relaxation are summarized in 
Table 1.
One can see that the relaxed bond length and the relaxed distance to the 
substrate vary from unrelaxed ones by as much as 0.2 \AA\ and 0.07\AA\ 
correspondently. 

\begin{table}
\label{tab1}
\caption{The bond length and vertical coordinates of Co$_4$ cluster.}
\begin{center}
\begin{tabular}{|l|c|c|c|}
\hline
          & ideal  & trial potentials     & ab initio     \\
\hline
 $r($\AA$)$    &2.56  &2.38      &  2.30     \\
\hline
 $z($\AA$)$    &0.00  &-0.07     & -0.05      \\
\hline
\end{tabular}
\end{center}
\end{table}

For the relaxed structure of the $Co_4$ cluster we 
perform the self-consistent KKR calculations in the full-potential 
approximation. The forces acting on Co adatoms and magnetic moments on 
atoms are obtained as the result of these calculations. We have found that 
relaxations(shortening of the bond length and decreasing of the distance 
to the substrate) has no strong effect on the spin magnetic 
moments of Co clusters in contrast to results for orbital moments and the 
magnetic 
anisotropy energy(Pick et al., 2003). This is because  the majority state  
of 
Co is practically filled. The effect of the substrate is mainly determined 
by the hybridization of the $sp$ Cu states with the $d$-states of the Co 
adatom. Our results show that the forces (vertical and horizontal) 
acting on Co atoms in Co$_4$ cluster after the relaxation with 
the trial potentials are 
not equal zero, see Fig 3.
\begin{figure}[htp]
\begin{center}
\includegraphics[width=14cm]{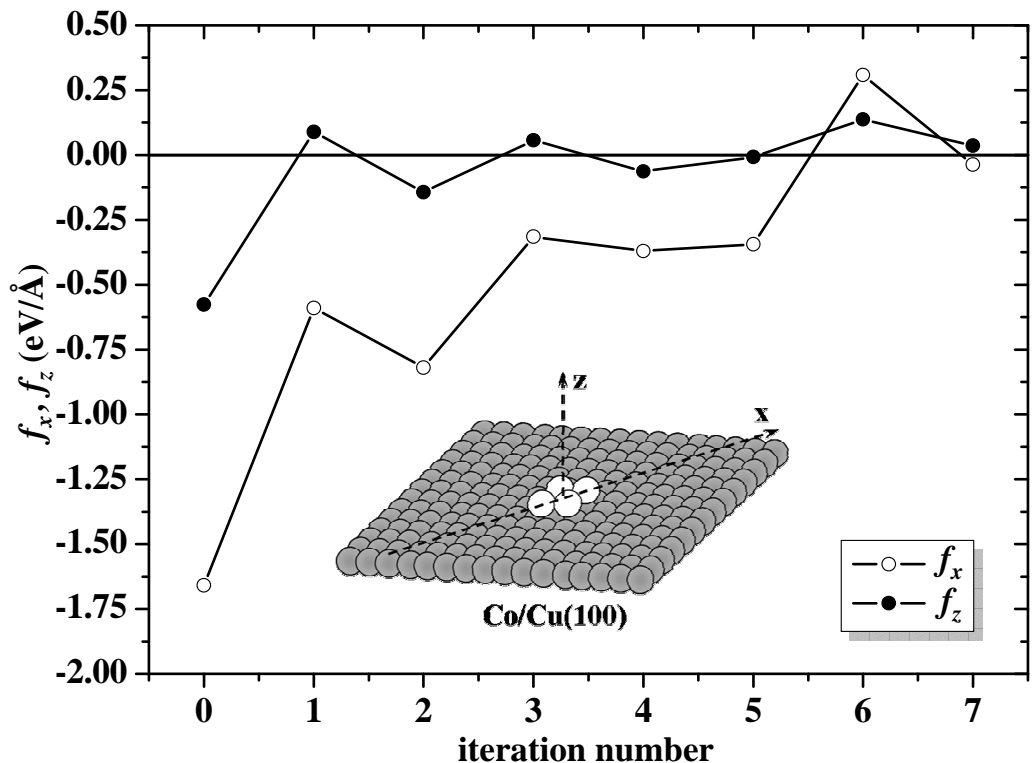}
\caption{ Forces acting on atoms in the $Co_4$ cluster during the
relaxation; variation in the vertical and horizontal forces is
presented.}\label{fig3}
\end{center}
\end{figure}
 We have fitted potentials for Co-Cu and Co-Co 
interactions to accurately reproduce these forces.
The cluster was again 
relaxed using a new potentials, and ab initio calculations of forces and 
magnetic moments were again  performed. We have found that only 7 
iterations(at each iteration the potentials are fitted to the ab initio 
forces) are necessary to find the equilibrium  structure of the cluster. 
Results presented in Fig.2  show that the interatomic potentials which 
nearly perfectly reproduce a fully ab initio results are very close to the 
trial potentials. It is seen  
that mainly the short-range repulsive part of the potential changes during the 
relaxation.    
Our analysis shows that electronic states and magnetic moments of clusters 
calculated for 
the relaxed  structure obtained with  the trial potentials (the first 
iteration, see Fig.4) are close to results for clusters relaxed fully ab 
initio(the last iteration, see Fig.4). 
\begin{figure}[htp]
\begin{center}
\includegraphics[height=16cm]{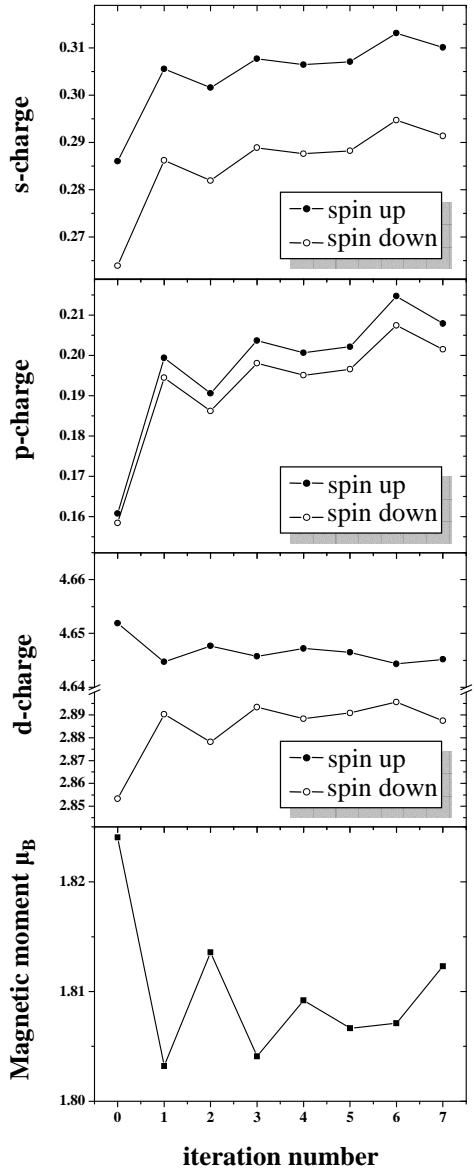}
\caption{Changes in the  charges in the WS cell and the local
magnetic moments on
atoms in the
$Co_4$ cluster during the relaxation.}\label{fig4}
\end{center}
\end{figure}
 The bond length in the Co$_4$ 
cluster 
after the ab initio relaxations is reduced by 0.08 \AA\ compared to 
results 
with the trial potentials. 

The above findings clearly demonstrate that
the interatomic potentials constructed by fitting  to an ab initio data 
pool  can be used to determine an   
equilibrium structure of supported clusters. Even after the 
first iteration (the calculation with the trial potentials) the 
geometrical structure, 
electronic and magnetic states are very close to the results of  
fully ab initio calculations.  We believe that, especially  in the case of 
large supported magnetic clusters, our approach for calculations of atomic  
relaxations and electronic states   can be very useful.
To the best of our knowledge such calculations have not been performed 
yet. 
   
We have also used our method to determine the relaxed structure and 
magnetic moments of Co$_2$, Co$_5$ and Co$_9$ clusters on the 
ideal Cu(001) surface. Similar to the calculation for the Co$_4$ , 
we needed only 6-7 iterations to find the equilibrium structure  for all 
 clusters. Our results for the bond lengths and magnetic moments are 
presented in Table 2. One can see that the trial potentials give a very 
good approximation for the geometrical and magnetic structure of clusters.

\begin{table}
\label{tab2}
\caption{The bond lengths, vertical coordinates and magnetic moments of
 Co$_2$, Co$_5$ and  Co$_9$ clusters.}

\begin{center}
\begin{tabular}{|c|c|c|c|c|}
\hline
 &          & ideal  & trial potentials     & ab initio     \\
\hline
       & $r($\AA$)$           & 2.56  &  2.37      &  2.29     \\
$Co_2$ & $z($\AA$)$           & 0.00  & -0.14      & -0.12     \\
       & $M (\mu B)$        & 1.89  &  1.86      &  1.86     \\
\hline
       & $r_{1,2} ($\AA$)$    & 2.56  &  2.39      &  2.36     \\
       & $z_1($\AA$)$         & 0.00  & -0.10      & -0.02     \\
$Co_5$ & $z_2($\AA$)$         & 0.00  & -0.03      & -0.06     \\
       & $M_1 (\mu B)$      & 1.73  &  1.64      &  1.68     \\
       & $M_2 (\mu B)$      & 1.85  &  1.85      &  1.83     \\
\hline
       & $r_{1,2} ($\AA$)$    & 2.56  &  2.40      &  2.38     \\
       & $r_{1,3} ($\AA$)$    & 3.61  &  3.43      &  3.37     \\
       & $z_1($\AA$)$         & 0.00  & -0.02      &  0.05     \\
       & $z_2($\AA$)$         & 0.00  & -0.04      &  0.00     \\
$Co_9$ & $z_3($\AA$)$         & 0.00  & -0.06      & -0.02     \\
       & $M_1 (\mu B)$      & 1.82  &  1.78      &  1.80     \\
       & $M_2 (\mu B)$      & 1.74  &  1.72      &  1.73     \\
       & $M_3 (\mu B)$      & 1.74  &  1.74      &  1.74     \\
\hline
\end{tabular}
\end{center}
\end{table}

\section {Summary}
By using the KKR Green's function method and an N-body potentials 
constructed by fitting to ab initio results, we have developed a new 
method for ab initio relaxations of magnetic supported clusters. This 
approach allows us to find the geometry, electronic and magnetic states in 
clusters fully self-consistently. 
We have demonstrated that only a few iterations should be performed in ab 
initio calculations to determine the structure and magnetic moments of 
small Co clusters on Cu(001).  

\section {Acknowledgements}
This work was supported by the Deutsche Forschungsgemeinschaft (DFG), 
Schwerpunktprogramm
"Cluster in Kontakt mit Oberfl\"achen: Elektronenstruktur und Magnetismus".
We thank A. N. Baranov for informative discussions.


\begin{thebibliography}{00}




\bibitem{c1} Brenner D.W. (2000), The art and science of an analytical 
potentials. Phys. Stat. Sol.(b), {\bf 217}, 23. 



\bibitem{c2} Levanov N.A., Stepanyuk V.S., Hergert W., Bazhanov D.I.,
Dederichs P., Katsnelson A., Massobrio C. (2000). Energetics of Co adatoms 
on the Cu(001) surface. Phys. Rev. B {\bf 61}, 2230. 


\bibitem{c3} Longo R., Stepanyuk V.S., Hergert W., Vega A., Gallego L.J., 
Kirschner J. (2004). Interface intermixing in metal heteroepitaxy on the 
atomic scale. Phys. Rev. B {\bf 69}, 073406. 

\bibitem{c4} Nayk S.K., Jena P., Stepanyuk V.S., Hergert W. (2001).
Effect of atomic relaxation on the magnetic moment of Fe, Co and Ni dimers 
supported on Cu(001). Surf. Sci.{\bf 491},219.



\bibitem{c5} Papanikolaou N., Zeller R., Dederichs P.H., Stefanou N. 
(1997). Lattice distortion in Cu-based dilute alloys: A first-principle 
study by the KKR Green-function method. Phys. Rev. B {\bf 55}, 4157 
 
\bibitem{c6} Pick S., Stepanyuk V.S., Baranov A.N., Hergert W., Bruno P. 
(2003). Effect of atomic relaxations on magnetic properties of adatoms and 
small clusters. Phys. Rev. B {\bf 68}, 104410.

\bibitem{c7} Stepanyuk V.S., Bazhanov D.I., Baranov A.N., Hergert W., 
Dederichs P.H., Kirschner J. (2000). Strain relief and island shape 
evolution in heteroepitaxial metal growth. Phys. Rev. B {\bf 62}, 15398.

\bibitem{c8} Stepanyuk V.S., Baranov A.N., Tsivlin D.V., Hergert W., 
Bruno 
P., Knorr N., Schneider M.A., Kern K. (2003). Quantum interference and 
long-range  adsorbate-adsorbate interactions. Phys. Rev. B {\bf 68}, 
205410.

\bibitem{c9} Stepanyuk V.S., Bruno P., Klavsyuk A.L., Baranov A.N., 
Hergert W., Saletsky A.M., Mertig I. (2004). Structure and quantum efects 
in atomic-sized contacts. Phys. Rev. B {\bf 69}, 033302.

\bibitem{c10} Tsivlin D.V., Stepanyuk V.S., Hergert W., Kirschner J. 
(2003).
Effect of mesoscopic relaxations on diffusion of Co adatoms on Cu(111).
Phys. Rev. B {\bf 68}, 205411.

\bibitem{c11} K.~Wildberger, V.S.~Stepanyuk, P.~Lang, R.~Zeller, and
P.H.~Dederichs. (1995). Magnetic nanostructures: 4d clusters on Ag(001). 
Phys. Rev. Lett. {\bf 75}, 509. 

\end{thebibliography}
\end{document}